# On static equilibrium and balance puzzler

Samrat Dey*[1], Ashish Paul[2], Dipankar Saikia[3], Deepjyoti Kalita[3], Anamika Debbarma[3], Shaheen Akhtar Wahab[3] and Saurabh Sarma[3]

[1]Department of Physics, DBCET, Assam Don Bosco University,
Guwahati -781017, INDIA.

[2]Department of Mathematics, DBCET, Assam Don Bosco University,
Guwahati -781017, INDIA.

[3]Department of Civil Engineering, DBCET, Assam Don Bosco University,
Guwahati -781017, INDIA.

Abstract

The principles of static equilibrium are of special interest to civil engineers. For a rigid body to be in static equilibrium the condition is that net force and net torque acting on the body should be zero. That clearly signifies that if equal weights are placed on either sides of a balance, the balance should be in equilibrium, even if its beam is not horizontal (we have considered the beam to be straight and have no thickness, an ideal case). Thus, although the weights are equal, they will appear different which is puzzling. This also shows that the concept of equilibrium is confusing, especially neutral equilibrium is confused to be stable equilibrium. The study not only throws more light on the concept of static equilibrium, but also clarifies that a structure need not be firm and steady even if it is in static equilibrium.

---

* Corresponding Author

1. Introduction

Static equilibrium represents a common situation in engineering practice, and the principles it involves are of special interest to civil engineers, architects, and mechanical engineers[1]. The condition for static equilibrium of a rigid body are that sum total of all the forces (**F**) and all the torques (**T**) acting on the body is equal to zero[2], i.e.,

$$\sum \boldsymbol{F} = 0 \quad \text{(i)}$$
$$\sum \boldsymbol{T} = 0. \quad \text{(ii)}$$

Using these equations, we analyse the static equilibrium condition of a balance, as an example of a rigid body. The reason for choosing the balance is that it is probably the simplest and most common rigid body in our day to day life where the application of Eqn. (i) and (ii), i.e., the conditions of equilibrium, can be observed with no trouble.

According to standard texts [3-6], a balance, a pivoted horizontal first order lever of equal length arms (connecting the axis of rotation to the point of force application[7] and thus, ideally having no width), called the beam, measures an unknown weight by the method of comparison based on the principle of moments; so in an ideal case, as both the arms can be horizontal (accordingly, they have an angle of 180° between them) and have negligible thickness, the fact that they are equal in length means that the beam is pivoted at its centre of gravity. So, placing two equal weights, $w_1$ (known) and $w_2$ (unknown), on the two arms, results in equal and opposite torques [thus, satisfying Eqn. (ii)] and the system gets balanced; Eqn. (i) is satisfied, because the total weight in the downward direction gets balanced by a reaction of equal magnitude at the pivot in the upward direction. Thus, we observe that the arms become horizontal and conclude that the unknown weight is equal to the known weight. However, it is easy to understand that when equal weights are placed, both Eqn. (i) and (ii) are satisfied even if the arms (the beam of the balance, *AB*) are not horizontal, as shown in Fig. A [although, in the figure, the beam (and thus, each arm) is shown to have some thickness, one can think it to have an infinitesimal thickness corresponding to a line connecting the points, *YOY´*, where *Y* and *Y´* are the points of application of weights and *O* is the point of suspension, which in this case is same as the centre of gravity, *S*]. This is because, two forces (the known and the unknown weights) and the perpendicular distances will still be same; taking *Y* & *Y´* to be same as *A* & *A´*, respectively, the perpendicular distances are *A´O* (= *AO* cos $\theta$) and *B´O* (= *BO* cos $\theta$), $\theta$ being the angle of inclination of the beam with the horizontal and *AO* & *BO* being the lengths of the arms, respectively; as *AO* = *BO*, *A´O* = *B´O*. Thus, as the beam remains in equilibrium in inclined position, the weights will appear different, although they are equal. This suggests that such a balance will not be able to measure weights. However, we know normal balances do measure weights. We shall revisit this problem in this article. This study not only sheds more light on the concept of static equilibrium but also clarifies that a structure need not be firm and steady even if it is in static equilibrium.

## 2. Analysis

Let us consider a balance (Fig. A), having a beam (*AB*), being suspended by *O* which divides the beam into two arms of equal length (*AO* and *OB*). If two equal weights are put on the end of the arms, as discussed above and shown in the Fig. A, the balance will be in equilibrium even if the beam is tilted making an angle $\theta$ with the horizontal. Thus, as argued earlier such a balance will not be able to measure weights. Now, the question is how in the case of a normal balance the beam remains horizontal when two equal weights are suspended by the ends and consequently, measures an unknown weight by the method of comparison.

In case of a normal balance, *S* and *O* are separated and *S* remains slightly below *O*, as shown in Fig. B, which is because the two arms are, basically, *XO* and *OX´*, where *X* and *X´* are points of application of weights; the arms have an angle (obtuse angle) between them, which is lesser than 180º (of the ideal case) and thus, *S* is lower than *O*. Thus, if the system is tilted on the left side (as shown in the Fig. B), the moment on the right side becomes more (because, angle becomes more) which tries to make the system horizontal. Another way of looking at the system is that, when equal weights are placed by the ends and the beam is tilted, the point *S* tries to again come back just below the point *O* and the beam becomes horizontal. This is because, if the beam is tilted a restoring couple (= $2.w.OS.\sin\theta$, where $w = w_1 = w_2$ is the weight placed on each end) comes into the picture, unless $\theta$ becomes equal to zero; in fact, the system is expected to execute damped simple harmonic motion, before $\theta$ becomes zero. When $\theta$ becomes zero the beam becomes horizontal and thus, the balance measures an unknown weight by the method of comparison.

If, however, *S* remains slightly above *O*, when equal weights are placed and the beam is tilted, the torque will rotate the balance further in the same direction in which it is tilted (Fig. C). Thus, measuring weights will, definitely, not be possible.

A balance with equal weights on either ends, in all the above three situations (if the beam is kept horizontal) satisfies the conditions of static equilibrium, in the first case it being in neutral equilibrium, in the second case it being in stable equilibrium and in the third case it being in unstable equilibrium. Thus, it is a common misconception that a balance (of the type, shown in Fig. A, in neutral equilibrium), as dictated by the definition (of being a pivoted horizontal equi-arm lever, consequently, meaning that it is pivoted at the centre of gravity, as discussed above), can compare equal weights by satisfying the condition of static equilibrium. It is only the balance in stable equilibrium that can measure weight. In some texts[3], however, it is accepted that in a balance *S* should be lower than *O*. However, we explicitly mention that the angle between the arms is basically obtuse angle and also, the three situations, corresponding to Fig. A, Fig. B and Fig. C, respectively, are discussed unambiguously.

From the above example we understand that just satisfying the condition of static equilibrium is not sufficient for a structure to be firm. A body may be in static equilibrium but may not be steady. The situation is puzzling especially in the case of neutral equilibrium. This is because, in neutral equilibrium, if the position of a structure is altered (even slightly) it continues to remain in that position (just like the ideal balance), unlike stable [unstable] equilibrium where a slight change will bring [take away] the body to [from] its initial position. Thus, a structure may apparently be steady but may not be actually so. The situation is similar to a heavy box kept on a frictionless surface. If resultant force on the body is zero, it may appear that the box will be firmly there at a fixed point. However, since the surface is frictionless even an infinitesimal force, tending to zero, may move the heavy box by a large distance.

Mathematically, however, the perception of the above three situations is very straight forward[5, 6]. If at a point, the tangent to (first derivative of) the potential curve (variation of potential energy as a function of distance) is zero, a body is at static equilibrium at that point. Now, if the point is a maximum, the body is at unstable equilibrium, if the point is a minimum, the body is at stable equilibrium and if it is neither a maximum, nor a minimum, the body is at neutral equilibrium.

### 3.    Conclusions

The present study clarifies the concept of static equilibrium, one of the most fundamental concepts required for civil engineers, architects, and mechanical engineers, by taking the example of a balance, possibly the simplest and most common rigid body where the application of the conditions of static equilibrium can be easily observed. We find that an ideal balance, being in neutral equilibrium when two equal weights are placed at the ends of its two arms, cannot measure weight by the method of comparison. This study clarifies that it is not simply the condition of static equilibrium that makes a structure firm and steady, the structure has to be in stable equilibrium. It also concludes that the situation is sometimes deceiving for neutral equilibrium (by taking the example of an ideal balance) where if the position of a structure is altered (even slightly) it continues to remain in that slightly altered position.

References:

1. http://www.electron.rmutphysics.com/physics/charud/scibook/Physics-for-Scientists-and-Engineers-Serway-Beichne%206edr-4/12%20-%20Static%20Equilibrium%20and%20Elasticity.pdf

2. http://civilengineering-notes.weebly.com/basic-concepts-in-statics-and-static-equilibrium.html

3. D. S. Mathur, Elements of Properties of Matter, Shyamlal Cheritable Trust, New Delhi, 1967.

4. Halliday, Resnick and Walker, Fundamentals of Physics, Wiley India, New Delhi, 2008.

5. H. D. Young and R. A. Freedman, Sears and Zemansky's University Physics, Pearson Education, 2006.

6. http://en.wikipedia.org/wiki/Weighing_scale

7. http://en.wikipedia.org/wiki/Torque

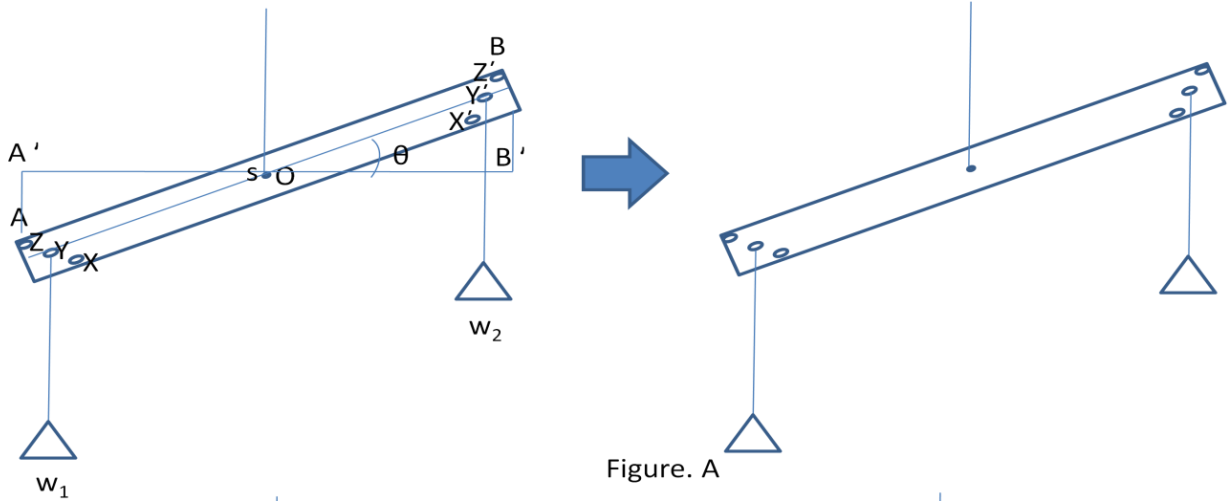
Figure. A

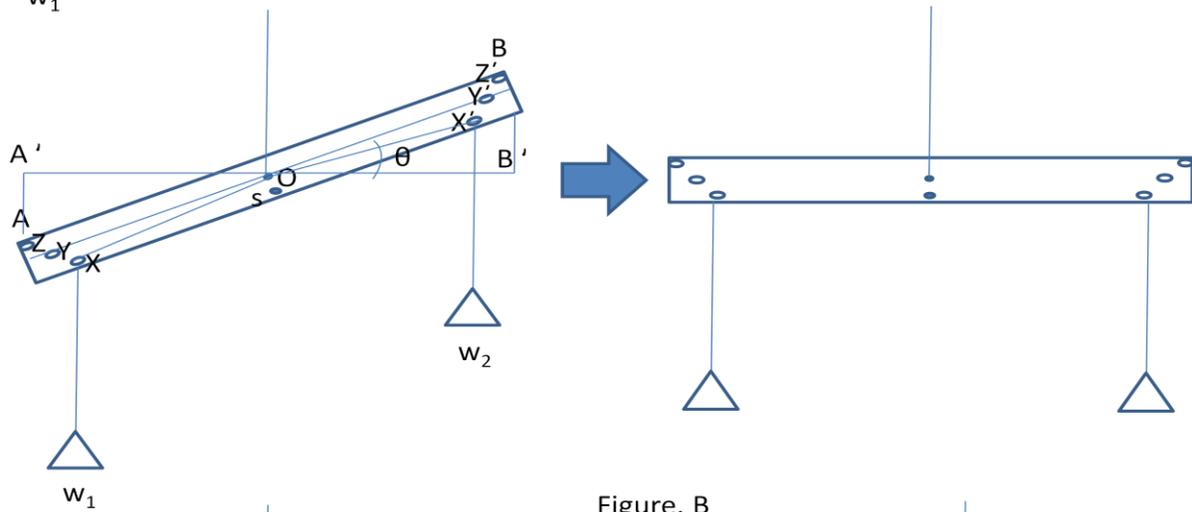
Figure. B

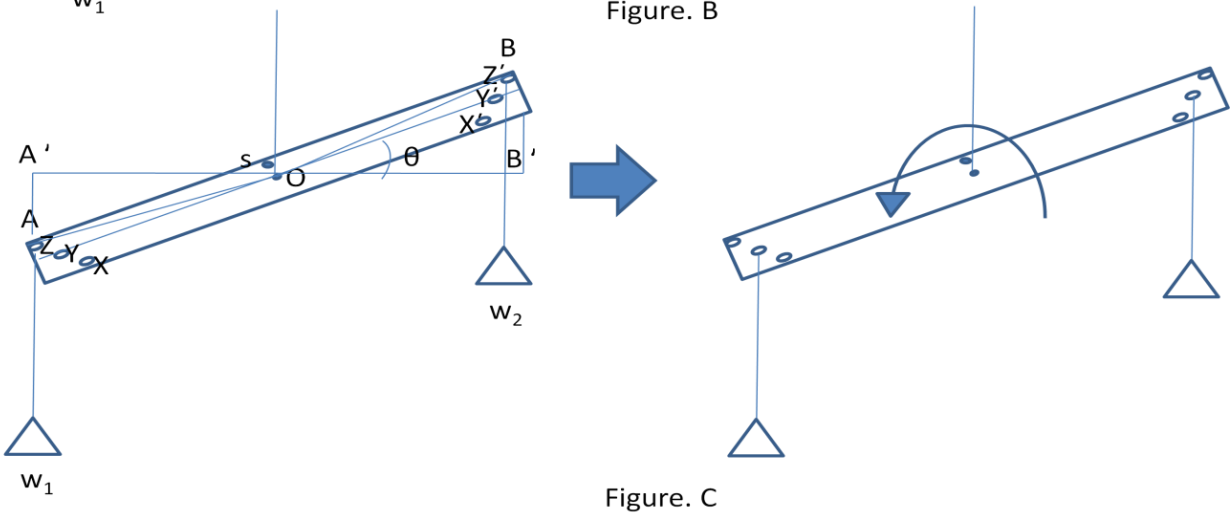
Figure. C

Figure A: A balance where point of suspension (*O*) coincides with centre of gravity (*S*). It continues to remain in the tilted position (even if $w_1 = w_2$). Figure B: A balance where the point of suspension lies slightly above the centre of suspension. It comes to horizontal position from the tilted position (if $w_1 = w_2$). Figure C: A balance where the point of suspension lies slightly below the centre of suspension. It moves further in the direction of tilted position (if $w_1 = w_2$).